\documentclass[12pt]{iopart}

\begin{document}

\title{The cosmic snap parameter in $f(R)$ gravity} 

\author{Nikodem J Pop\l awski}

\address{Department of Physics, Indiana University, 
727 East Third Street, Bloomington, IN 47405, USA}
\ead{nipoplaw@indiana.edu}

\begin{abstract}
We derive the expression for the snap parameter in $f(R)$ gravity. 
We use the Palatini variational principle to obtain the field equations
and regard the Einstein conformal frame as physical.
We predict the present-day value of the snap parameter for the particular case $f(R)=R-\mbox{const}/R$, which is the simplest $f(R)$ model explaining the current acceleration of the universe.
\end{abstract}


\maketitle

\section{Introduction}

$f(R)$ gravity models, in which the gravitational Lagrangian is a function of 
the curvature scalar $R$~\cite{fR1,fR2}, have recently attracted a lot of interest.
They explain how the current cosmic acceleration originates from the
addition of a term $R^{-1}$ to the Einstein--Hilbert Lagrangian
$R$~\cite{acc}.
As in general relativity, $f(R)$ gravity theories obtain the field equations
by varying the total action for both the field and matter,
and equaling this variation to zero.
Here we use the {\it metric--affine} (Palatini) variational principle,
according to which the metric $g_{\mu\nu}$ and the affine connection 
$\Gamma^{\,\,\rho}_{\mu\,\nu}$ are considered as 
geometrically independent quantities, and the action is varied 
with respect to both of them~\cite{Schr,OK,Niko1,SL}.
The standard approach is the {\it metric} (Einstein--Hilbert) variational principle, 
according to which the action is varied with respect to the metric 
tensor, and the affine connection coefficients are
the Christoffel symbols of $g_{\mu\nu}$~\cite{LL}.
Both approaches give the same result 
only if we use the standard Einstein--Hilbert action~\cite{Schr}.
The field equations in the Palatini formalism are second-order 
differential equations, while for metric theories they are 
fourth-order~\cite{Mag}.
Another remarkable property of the metric--affine approach is that the 
field equations in vacuum reduce to the standard Einstein equations of 
general relativity with a cosmological constant~\cite{Mag,cosm}.

One can show that $f(R)$ theories of gravitation 
are conformally equivalent to the Einstein theory
of the gravitational field interacting with additional matter 
fields, if the action for matter does not depend on the connection~\cite{SL,Mag,eq}. 
This can be done by means of a Legendre 
transformation, which replaces an $f(R)$ Lagrangian with a
Helmholtz Lagrangian~\cite{Mag,Hel}.
For $f(R)$ gravity, the scalar degree of freedom 
due to the occurrence of nonlinear second-order terms in the Lagrangian 
is transformed into an auxiliary scalar field $\phi$~\cite{eq}.
The set of variables $(g_{\mu\nu},\,\phi)$ is commonly called the 
{\it Jordan conformal frame}. 
In the Jordan frame, the connection is metric incompatible unless $f(R)=R$.
The compatibility can be restored by a certain conformal 
transformation of the metric: $g_{\mu\nu}\rightarrow 
h_{\mu\nu}=f'(R)g_{\mu\nu}$~\cite{conf}.
The new set $(h_{\mu\nu},\,\phi)$ is called the {\it Einstein conformal 
frame}, and we will regard the metric in this frame as physical.
Although both frames are equivalent mathematically, they are {\it not}  
equivalent physically~\cite{EJ2}, and
the interpretation of cosmological observations can drastically 
change depending on the adopted frame~\cite{EJ1}.

$f(R)$ gravity models in both the metric and metric--affine formalism
have been compared with cosmological observations by
several authors~\cite{obs,SO,repro,uniq,Niko2}.
The problem of viability of these models at smaller scales, namely their compatibility with solar system observations, is a subject of recent debate (\cite{Far} and references therein).
There are also limits on these models arising from particle physics \cite{Flan}, matter instability~\cite{inst} and violation of the equivalence principle~\cite{EJ2,Olmo}.
Current SNIa observations provide the data on the time evolution of the
deceleration parameter $q$ in the form of $q=q(z)$, where $z$ is the
redshift~\cite{Gold}.
The extraction of the information from these data depends, however, on
assumed parametrization of $q(z)$~\cite{jerk2}.
For small values of $z$ such a dependence can be linear,
$q(z)=q_0+q_1 z$~\cite{Gold}, but its validity should fail at $z\sim 1$.
A convenient method to describe models close to $\Lambda CDM$ is based on
the cosmic jerk parameter $j$, a dimensionless third derivative of the scale
factor with respect to the cosmic time~\cite{jerk1,snap}.
A deceleration-to-acceleration transition occurs for models with
a positive value of $j_0$ and negative $q_0$.ff
The flat $\Lambda CDM$ models have a constant jerk $j=1$.

Recently, we showed~\cite{Niko3} that the predictions for the current value
of the jerk parameter for the particular case $f(R)=R-\frac{\alpha^2}{3R}$,
which is the simplest way of introducing the cosmological term in $f(R)$
gravity~\cite{acc,Niko2}, agree with the SNLS SNIa~\cite{SNLS} and x-ray galaxy cluster
distance~\cite{jerk2} data, but do not with the SNIa gold sample data~\cite{Gold}.
Moreover, the predicted value of the deceleration parameter in this model
agrees with all three data sets~\cite{Niko3}.
Therefore $f(R)$ models based on the Palatini variational principle and
formulated in the Einstein frame,
including the case $f(R)=R-\frac{\alpha^2}{3R}$, provide a possible
explanation of the current cosmic acceleration.
More constraints on these models are likely to come from nondimensional
parameters containing higher derivatives of the scale factor, such as the
snap parameter $s=\frac{\ddot{\ddot{a}}}{aH^4}$~\cite{snap}.

In this work, we find the general expression for the snap parameter
in $f(R)$ gravity, assuming that the universe is flat, homogeneous,
isotropic and pressureless.
We use the field equations derived from the Palatini variational principle.
We assume that matter is minimally coupled to the metric tensor in the
Jordan frame, and then transform to the Einstein frame which we consider physical~\cite{Niko1}
and in which this coupling is non-minimal~\cite{OK}.
Since the question of which frame is physical and in which frame
the matter--metric coupling is minimal should be ultimately
answered by experiment or observation, the presented model should be treated as phenomenological. 
Anticipating cosmological measurements, we predict the
current value of the snap parameter for the case
$f(R)=R-\frac{\alpha^2}{3R}$.

\section{The field equations in $f(R)$ gravity}

The action for $f(R)$ gravity in the original (Jordan) frame 
with the metric $\tilde{g}_{\mu\nu}$ is given by~\cite{Niko1}
\begin{equation}
S_J=-\frac{1}{2\kappa c}\int d^4 
x\bigl[\sqrt{-\tilde{g}}f(\tilde{R})\bigr]+S_m(\tilde{g}_{\mu\nu},\psi).
\label{action1}
\end{equation}
Here, $\sqrt{-\tilde{g}}f(\tilde{R})$ is a Lagrangian density that depends 
on the curvature scalar:
$\tilde{R}=R_{\mu\nu}(\Gamma^{\,\,\lambda}_{\rho\,\sigma})\tilde{g}^{\mu\nu}$, 
$S_m$ is the action for matter represented 
symbolically by $\psi$ and independent of the connection, 
and $\kappa=\frac{8\pi G}{c^4}$. 
Tildes indicate quantities calculated in the Jordan frame.

Varying the action $S_J$ with respect to $\tilde{g}_{\mu\nu}$ 
yields the field equations:
\begin{equation}
f'(\tilde{R})R_{\mu\nu}-\frac{1}{2}f(\tilde{R})\tilde{g}_{\mu\nu}=\kappa 
T_{\mu\nu},
\label{field1}
\end{equation} 
where the dynamical energy--momentum tensor of matter, $T_{\mu\nu}$, is generated 
by the Jordan metric tensor~\cite{OK,Niko1}:
\begin{equation}
\delta S_m=\frac{1}{2c}\int d^4 x\sqrt{-\tilde{g}}\,T_{\mu\nu}\delta\tilde{g}^{\mu\nu},
\label{EMT1}
\end{equation} 
and the prime denotes the derivative of a function with respect to its variable.
From vanishing of the variation of $S_J$ with
respect to the connection $\Gamma^{\,\,\rho}_{\mu\,\nu}$
it follows that this connection is given by the
Christoffel symbols of the conformally transformed metric~\cite{eq,conf}
\begin{equation}
g_{\mu\nu}=f'(\tilde{R})\tilde{g}_{\mu\nu}.
\label{conf}
\end{equation}
The metric $g_{\mu\nu}$ defines the Einstein frame, in which the connection is metric compatible.

The action~(\ref{action1}) is dynamically equivalent
to the following Helmholtz action~\cite{Niko1,Mag,eq}:
\begin{equation}
S_H=-\frac{1}{2\kappa c}\int d^4 x\sqrt{-\tilde{g}}\bigl[f(\phi(p))+p(\tilde{R}-\phi(p))\bigr]+S_m(\tilde{g}_{\mu\nu},\psi),
\label{action2}
\end{equation}
where $p$ is a new scalar variable.
The function $\phi(p)$ is determined by
\begin{equation}
\frac{\partial 
f(\tilde{R})}{\partial\tilde{R}}\bigg{\vert}_{\tilde{R}=\phi(p)}=p.
\label{phi}
\end{equation}
From equations~(\ref{conf}) and~(\ref{phi}) it follows that
\begin{equation}
\phi=Rf'(\phi),
\label{resc}
\end{equation}
where $R=R_{\mu\nu}(\Gamma^{\,\,\lambda}_{\rho\,\sigma})g^{\mu\nu}$ is
the Riemannian curvature scalar of the metric $g_{\mu\nu}$.

In the Einstein frame, the action~(\ref{action2}) becomes the standard 
Einstein--Hilbert action of general relativity with an additional
scalar field:
\begin{equation}
S_E=-\frac{1}{2\kappa c}\int d^4 x\sqrt{-g}\bigl[R-p^{-1}\phi(p)+p^{-2}f(\phi(p))\bigr]+S_m(p^{-1}g_{\mu\nu},\psi).
\label{action3}
\end{equation}
Choosing $\phi$ (which is equal to the curvature scalar $\tilde{R}$ in the Jordan frame)
as the scalar variable leads to
\begin{equation}
S_E=-\frac{1}{2\kappa c}\int d^4 x\sqrt{-g}\bigl[R-2V(\phi)\bigr]+S_m([f'(\phi)]^{-1}g_{\mu\nu},\psi),
\label{action4}
\end{equation}
where $V(\phi)$ is the effective potential:
\begin{equation}
V(\phi)=\frac{\phi f'(\phi)-f(\phi)}{2[f'(\phi)]^2}.
\label{pot}
\end{equation}

Varying the action~(\ref{action4}) with respect to 
$g_{\mu\nu}$ yields the equations of the gravitational field in
the Einstein frame~\cite{OK,Niko1}: 
\begin{equation}
R_{\mu\nu}-\frac{1}{2}Rg_{\mu\nu}=\frac{\kappa 
T_{\mu\nu}}{f'(\phi)}-V(\phi)g_{\mu\nu},
\label{EOF1}
\end{equation}
while the variation with respect to $\phi$ reproduces~(\ref{resc}).
Equations~(\ref{resc}) and~(\ref{EOF1}) give
\begin{equation}
\phi f'(\phi)-2f(\phi)=\kappa Tf'(\phi),
\label{struc2}
\end{equation}
from which we obtain $\phi=\phi(T)$.
Substituting $\phi$ into the field equations~(\ref{EOF1}) leads
to a relation between the Ricci tensor and the energy--momentum tensor,
\begin{equation}
R_{\mu\nu}-\frac{1}{2}Rg_{\mu\nu}=\kappa_r(T)T_{\mu\nu}+\Lambda(T)g_{\mu\nu},
\label{EOF2}
\end{equation}
with a running gravitational coupling
$\kappa_r(T)=\kappa[f'(\phi(T))]^{-1}$
and a variable cosmological term $\Lambda(T)=-V(\phi(T))$:
\begin{equation}
\Lambda(\phi)=\frac{f(\phi)-\phi f'(\phi)}{2[f'(\phi)]^2}.
\label{cc}
\end{equation}
Such a relation is in general nonlinear and depends on the form of the
function $f(R)$.

The Bianchi identity applied to equation~(\ref{EOF1}) gives
\begin{equation}
T_{\mu\nu}^{\phantom{\mu\nu};\nu}=\phi^{,\nu}f''(\phi)\Bigl(\frac{T_{\mu\nu}}{f'(\phi)}+\frac{[2f(\phi)-\phi f'(\phi)]g_{\mu\nu}}{2\kappa[f'(\phi)]^2}\Bigr).
\label{conserv}
\end{equation} 
This relation means that the energy--momentum tensor in the Einstein frame is
not covariantly conserved, unless $f(R)=R$ or $T=0$~\cite{Niko2}.
We can write the field equation~(\ref{EOF1}) as
\begin{equation}
R_{\mu\nu}-\frac{1}{2}Rg_{\mu\nu}=\kappa(T_{\mu\nu}^m+T_{\mu\nu}^{\Lambda}),
\label{EOF3}
\end{equation}
where $T_{\mu\nu}^m=T_{\mu\nu}$.
This defines the dark energy--momentum tensor,
\begin{equation}
T_{\mu\nu}^{\Lambda}=\frac{\Lambda(\phi)}{\kappa}g_{\mu\nu}+\frac{1-f'(\phi)}{f'(\phi)}T_{\mu\nu}.
\label{darkt}
\end{equation}
From equation~(\ref{EOF3}) it follows that matter and dark energy
form together a system that has a conserved 4-momentum.
Consequently, in the Palatini $f(R)$ gravity formulated in the Einstein frame,
matter and dark energy {\it interact} (\cite{Niko4} and references therein).
This interaction may be responsible for the observed large entropy of the universe.

We assumed that matter is minimally coupled to the metric tensor in the
Jordan frame.
Then we transformed to the Einstein frame, in which this coupling becomes non-minimal,
and assumed that this frame is physical, motivated by the fact that the connection is
metric compatible in this frame.
Such a construction is completely phenomenological.
However, if we consider the Einstein frame from the very beginning and define
the energy--momentum tensor generated by the metric tensor $g_{\mu\nu}$ as the true 
energy--momentum tensor for matter (minimal coupling in the Einstein frame),
the resulting action, instead of equation~(\ref{action4}), is
\begin{equation}
S_E=-\frac{1}{2\kappa c}\int d^4 x\sqrt{-g}\bigl[R-2V(\phi)\bigr]+S_m(g_{\mu\nu},\psi).
\label{action5}
\end{equation}
Consequently, the field $\phi$ does not couple to anything and we arrive at
general relativity with the cosmological constant, which is not interesting 
from a modified gravity perspective~\cite{OK,Niko1}.

\section{The snap parameter in $f(R)$ gravity}

The snap parameter in cosmology is defined as~\cite{snap}
\begin{equation}
s=\frac{\ddot{\ddot{a}}}{aH^4},
\label{sn1}
\end{equation}
where $a$ is the cosmic scale factor, $H$ is the Hubble parameter, and the
dot denotes differentiation with respect to the cosmic time.
This parameter appears in the fourth-order term of the Taylor expansion of the
scale factor around $a_0$:
\begin{eqnarray}
& & \frac{a(t)}{a_0}=1+H_0(t-t_0)-\frac{1}{2}q_0H^2_0(t-t_0)^2+\frac{1}{6}j_0H^3_0(t-t_0)^3 \nonumber \\
& & +\frac{1}{24}s_0H^4_0(t-t_0)^4+\,O[(t-t_0)^5],
\label{exp}
\end{eqnarray}
where the subscript $0$ denotes the present-day value.
We can rewrite equation~(\ref{sn1}) as
\begin{equation}
s=\frac{\dot{j}}{H}-j(2+3q),
\label{sn2}
\end{equation}
where $q$ is the deceleration parameter and $j$ is the jerk parameter.
For the flat $\Lambda CDM$ model $s=-(2+3q)$ since $j=1$~\cite{jerk2,Niko3},
and the departure of the quantity $ds/dq$ from $-3$ measures how the evolution of the universe
deviates from the $\Lambda CDM$ dynamics.

From the gravitational field equations~(\ref{EOF1}) applied to a flat
Robertson--Walker universe filled with dust
we can derive the $\phi$-dependence of the Hubble parameter~\cite{Niko1}
\begin{equation}
H(\phi)=\frac{c}{f'(\phi)}\sqrt{\frac{\phi f'(\phi)-3f(\phi)}{6}},
\label{Hub1}
\end{equation}
the deceleration parameter~\cite{Niko2}
\begin{equation}
q(\phi)=\frac{2\phi f'(\phi)-3f(\phi)}{\phi f'(\phi)-3f(\phi)},
\label{dec1}
\end{equation}
and the jerk parameter~\cite{Niko3}
\begin{eqnarray}
& & j(\phi)=[2\phi^2 f'^4+10\phi^3 f'^3 f''-75\phi^2 f'^2 ff''-12\phi ff'^3+18f^2 f'^2-162f^3 f'' \nonumber \\
& & +189\phi f^2 f'f'']\times[(\phi f'-3f)^2(2f'^2+\phi f'f''-6ff'')]^{-1}.
\label{jer1}
\end{eqnarray}
Th prime denotes the differentiation with respect to $\phi$.
We also have the expression for the time dependence of $\phi$~\cite{Niko1}:
\begin{equation}
\dot{\phi}=\frac{\sqrt{6}c(\phi f'-2f)\sqrt{\phi f'-3f}}{2f'^2+\phi 
f'f''-6ff''}.
\label{phidot2}
\end{equation}

For the $\phi$-derivative of the jerk parameter we obtain a quite
complicated expression:
\begin{eqnarray}
& & j'=[(\phi f'-3f)(2f'^2+\phi f'f''-6ff'')(30\phi^3 f'^2 f''^2+10\phi^3 f'^3 f''' \nonumber \\
& & -150\phi^2 ff'f''^2-37\phi^2 f'^3 f''-75\phi^2 ff'^2 f'''-8\phi f'^4+24ff'^3+189\phi f^2 f''^2 \nonumber \\
& & +189\phi f^2 f'f'''+192\phi ff'^2 f''-162f^3 f'''-267f^2 f'f'') \nonumber \\
& & -(2\phi^2 f'^4+10\phi^3 f'^3 f''-75\phi^2 f'^2 ff''-12\phi ff'^3+18f^2 f'^2+189\phi f^2 f'f'' \nonumber \\
& & -162f^3 f'')\times(3\phi^2 f'f''^2-15\phi ff''^2-8f'^3+27ff'f''-\phi f'^2f''+\phi^2 f'^2 f''' \nonumber \\
& & -9\phi ff'f'''+18f^2 f''')]\times[(\phi f'-3f)^3(2f'^2+\phi f'f''-6ff'')^2]^{-1}.
\label{sn3}
\end{eqnarray}
Combining equations~(\ref{sn2}--\ref{phidot2}) and using
$\dot{j}=\dot{\phi}j'(\phi)$ lead to
\begin{equation}
s=j'\frac{6f'(\phi f'-2f)}{(2f'^2+\phi f'f''-6ff'')}-j\frac{8\phi f'-15f}{\phi f'-3f}.
\label{sn4}
\end{equation}
Putting here $j$ from equation(\ref{jer1}) and $j'$ from equation~(\ref{sn3}) gives the final
expression for the snap parameter in $f(R)$ gravity as a function of
$\phi$, $f(\phi)$, $f'(\phi)$, $f''(\phi)$, and $f'''(\phi)$,
which we do not write explicitly.

We now examine the case $f(R)=R-\frac{\alpha^2}{3R}$, where $\alpha$ is
a constant, which is a possible
explanation of the current cosmic acceleration~\cite{acc}.
In this model, the present-day value of $\phi$ is $\phi_0=(-1.05\pm0.01)\alpha$,
where $\alpha=(7.35^{+1.12}_{-1.17})\times10^{-52}m^{-2}$~\cite{Niko2}.
We do not need to know the exact value of $\alpha$ since it does not affect
nondimensional cosmological parameters.
Substituting $\phi_0$ into equations (\ref{jer1}), (\ref{sn3}) and (\ref{sn4}) gives
the present-day value of the cosmic snap parameter:\footnote{
The predicted value for the current cosmic jerk parameter found in~\cite{Niko3} is $j_0=1.01^{+0.08}_{-0.21}$. Here, we recalculated this value and obtained $j_0=1.01\pm0.01$, which differs from the former by the precision errors. This correction does not change the conclusions of~\cite{Niko3}.}
\begin{equation}
s_0=-0.22^{+0.21}_{-0.19}.
\label{snval1}
\end{equation}
In the $f(R)=R-\frac{\alpha^2}{3R}$ model, the deceleration-to-acceleration
transition occurred at $\phi_t=-\sqrt{5/3}\alpha$~\cite{Niko2}.
Consequently, we find the snap parameter at this moment:
\begin{equation}
s_t=-2.68.
\label{snval2}
\end{equation}
This value shows that the snap parameter in $f(R)$ gravity changes
significantly between the deceleration-to-acceleration transition and now,
which is clear from equation~(\ref{sn2}) and the fact that the deceleration parameter changes in this
period of time from 0 to the predicted value $q_0=-0.67^{+0.06}_{-0.03}$~\cite{Niko2}.
For the flat $\Lambda CDM$ model, the snap parameter increases from
$s=-7/2$ for the matter epoch, through $s=-2$ at this transition, to
the asymptotic de Sitter value $s=1$, indicating the difference between
the $f(R)$ and $\Lambda CDM$ predictions for $s_t$.

Lastly, we show the role of the energy conditions~\cite{Wald} in
metric--affine $f(R)$ gravity models.
For a pressureless universe, these conditions reduce to the inequality
\begin{equation}
\epsilon\geq0,
\label{cond1}
\end{equation}
where $\epsilon=T$ is the energy density of matter.
From equations (\ref{struc2}), (\ref{Hub1}) and (\ref{dec1}) we obtain
\begin{equation}
\epsilon=\frac{2H^2(1+q)f'(\phi)}{\kappa c^2}.
\label{cond2}
\end{equation}
Since $q>-1$ (it reaches $-1$ asymptotically~\cite{Niko2}) and $f'(\phi)>0$ (this condition
assures that the conformal transformation from the Jordan to the Einstein
frame does not change the signature of the metric tensor),
formula~(\ref{cond1}) is satisfied.
Therefore, the energy conditions do not impose additional constraints on
Palatini $f(R)$ gravity models.
For metric $f(R)$ models, these conditions lead to constraints containing
the jerk and snap parameter~\cite{ener}.\footnote{
The jerk and snap parameters do not appear in equation~(\ref{cond2}) and the
energy conditions because the field equations are second order, as opposed
to the fourth-order field equations in the metric formalism.}

\section{Summary}

We derived the expression for the cosmic snap parameter in
$f(R)$ gravity formulated in the Einstein conformal frame.
We used the Palatini variational principle to obtain the field equations
and apply them to a flat, homogeneous, and isotropic universe filled with dust.
We considered the particular case $f(R)=R-\frac{\alpha^2}{3R}$, which is
the simplest $f(R)$ model explaining the current cosmic acceleration,
and for which the predicted present-day values of the deceleration and jerk parameters
are quite consistent with cosmological data.
For the present-day value of the snap parameter, we predict $s_0=-0.22^{+0.21}_{-0.19}$.

Expanding the scale factor to fourth order with respect to time (equation~(\ref{exp})) is physically meaningful, since cosmological data already allow
to measure the third-order term: the jerk parameter~\cite{jerk2}.
The snap parameter may be important for observations involving redshifts $z\sim1$ and higher, where 
expansion of cosmological quantities in powers of $z$ cannot be limited only to linear and quadratic terms.
Therefore this paper is not only a formal mathematical exercise,
but also provides physically measureable constraints on Palatini $f(R)$ gravity.

A cosmological sequence of matter dominance, deceleration-to-acceleration transition and acceleration era
may always emerge as cosmological solutions of $f(R)$ gravity~\cite{SO}.
We showed in~\cite{Niko4} that, in Palatini $f(R)$ gravity, the deviation of the growth of the cosmic scale factor in the matter era from the standard law $a(t)\sim t^{2/3}$ is small, which is consistent with WMAP
cosmological data~\cite{WMAP}.
On the other hand, metric $f(R)$ gravity models with a power of $R$ dominant at large or small $R$ yield
the law $a(t)\sim t^{1/2}$ which is ruled out by cosmological observations~\cite{viab}.
Therefore $f(R)$ gravity in the Palatini variational formalism is a viable theory of gravitation that explains
the current cosmic acceleration.

It is possible to find an $f(R)$ action without cosmological constant which exactly reproduces the behavior of the Einstein--Hilbert action with cosmological constant, i.e. the expansion history of the universe does not uniquely determine the form of the gravitational action~\cite{repro,uniq}.
Moreover, the background expansion alone cannot distinguish
between different choices of $f(R)$ and one must study cosmological perturbations in order to determine
which choice is physical~\cite{uniq,pert}.
Measurements of higher derivatives of the scale factor, including the snap parameter,
will probably constitute a robust part of these studies.

\end{document}